\RequirePackage{fix-cm}
\documentclass[twocolumn,epjc3]{svjour3}  
\smartqed  
\RequirePackage{graphicx}
\usepackage{multicol}
\RequirePackage{float}
\RequirePackage{hyperref}
\RequirePackage{amsmath}
\RequirePackage{amssymb}
\usepackage{widetext}
\RequirePackage{mathtools}  
\usepackage{caption}
\usepackage{subcaption}    
\journalname{Eur. Phys. J. C}
\begin{document}

\title{Baryogenesis in $f(Q,\mathcal{ T})$ Gravity}


\author{Snehasish Bhattacharjee\thanksref{e1,addr1}
                \and
        P.K. Sahoo\thanksref{e2,addr2} 
      }

\thankstext{e1}{e-mail: snehasish.bhattacharjee.666@gmail.com}
\thankstext{e2}{e-mail: pksahoo@hyderabad.bits-pilani.ac.in}


\institute{Department of Astronomy, Osmania University, Hyderabad-500007,
India \label{addr1}
           \and
           Department of Mathematics, Birla Institute of
Technology and Science-Pilani, Hyderabad Campus, Hyderabad-500078,
India \label{addr2}
     }

\date{Received: 12 February 2020 / Accepted: 13 March 2020}

\maketitle

\begin{abstract}
The article communicates exploration of gravitational baryogenesis in presence of $f(Q,\mathcal{ T})$ gravity where $Q$ denote the nonmetricity and $\mathcal{ T}$ the trace of the energy momentum tensor. We study various baryogenesis interactions proportional to $\dot{Q}$ and $\dot{Q}f_{Q}$ for the $f(Q,\mathcal{ T})$ gravity model $f(Q,\mathcal{ T})=\alpha Q ^{n+1} + \beta \mathcal{ T}$, where $\alpha$, $\beta$ and $n$ are model parameters. Additionally we report the viable parameter spaces for which an observationally consistent baryon-to-entropy can be generated. Our results indicate that $f(Q,\mathcal{ T})$ gravity can contribute significantly and consistently to the phenomenon of gravitational baryognesis.

\keywords{ $f(Q,T)$ Gravity \and Modified Gravity \and Baryogenesis \and Early Universe \and Equation of State}
 \PACS{04.50.kd.}
\end{abstract}

\section{Introduction}\label{sec:int}
Our universe favors matter over antimatter for some mysterious reasons. Observations from Cosmic Microwave Background \cite{fg2}, coupled with successful predictions from the Big Bang Nucleosynthesis \cite{fg1}, recommend an overwhelming supremacy of matter over antimatter. \\
Cosmological theories that aim at resolving this fundamental issue falls under the domain of Baryogenesis. Theories such as GUT baryogenesis, Thermal Baryogenesis, Affleck-Dine Baryogenesis, Electroweak Baryogenesis, Black hole evaporation Baryogenesis and Spontaneous Baryogenesis propose interactions which goes beyond the standard model to explain this profound dominance of matter in the universe \cite{frt9}. These mechanisms were further developed in \cite{frt40}\\
Gravitational Baryogenesis is one such theory proposed in \cite{dav} and further developed and extended to many modified gravity theories \cite{fg4}. This particular theory employs one of the Sakharov criterion \cite{fg9} which assures a baryon- antibaryon asymmetry from the existence of a CP- violating interaction, which reads 
\begin{equation}\label{11}
\frac{1}{M_{*}^{2}}\int \sqrt{g}(\partial_{i}R)J^{i} d^{4}x 
\end{equation}       
where $M_{*}$ is the mass parameter of the underlying effective theory, $g$, $J^{i}$ and $R$ denote respectively the metric scalar, baryon current and Ricci scalar. Hence, for a flat FLRW background, the baryon to entropy ratio $\eta_{B}/s$ is proportional to time derivative of Ricci scalar $\dot{R}$. For a radiation dominated universe with EoS parameter $\omega = p / \rho = 1/3$, the net baryon asymmetry produced by \eqref{11} is zero.\\ 
The paper aims at investigating gravitational Baryogenesis through other curvature invariants and specifically through the nonmetricity $Q$. For the $f(Q,\mathcal{ T})$ gravity, the CP-violating interaction is given by 
\begin{equation}\label{12}
\frac{1}{M_{*}^{2}}\int \sqrt{g}(\partial_{i}(Q+\mathcal{ T}))J^{i} d^{4}x 
\end{equation}
where $\mathcal{ T}$ denote the trace of energy momentum tensor and the nonmetricity $Q$ is defined as \cite{fqt} 
\begin{equation}\label{2}
Q =6\frac{H^{2}}{N^{2}}
\end{equation}
where $H(t)$ represents Hubble parameter and $N(t)$ the lapse function. A remarkable difference between \eqref{11} with \eqref{12} is that the latter yields a nonzero baryon asymmetry even for a radiation dominated universe ($\omega = 1/3$). We shall investigate here in detail the baryogenesis terms proportional to $\partial_{i}Q$ and $\partial_{i}f(Q)$ and compare our results with cosmological observations. The paper is organized as follows: In Section \ref{sec:fqt} we provide a summary of $f(Q,\mathcal{ T})$ gravity and obtain the field equations. In Section \ref{sec:baryo} we explain in detail the gravitational baryogenesis in $f(Q,\mathcal{ T})$ gravity and infer the viability of a $f(Q,\mathcal{ T})$ gravity model in producing observationally acceptable baryon to entropy ratio and finally in Section \ref{sec:con} we present our conclusions.

\section{Field Equations in $f(Q,\mathcal{ T})$ Gravity}\label{sec:fqt}

The action in $f(Q,\mathcal{ T})$ gravity is given as \cite{fqt}
\begin{equation} \label{1}
\mathcal{S} = \int \left[  \frac{1}{16 \pi} f(Q,\mathcal{ T}) + \mathcal{L}_{M} \right] \sqrt{-g}d^{4}x
\end{equation} 
where $g\equiv det (g_{ij})$ denote the metric scalar. \\
Variation of action \eqref{1} with respect to metric tensor components yields the field equations in $f(Q,\mathcal{ T})$ gravity as \cite{fqt}
\begin{widetext}
\begin{equation}\label{3}
8 \pi \mathcal{ T}_{ij} = -\frac{2}{\sqrt{-g}}\bigtriangledown_{\alpha} \left( f_{Q}\sqrt{-g} P^{\alpha}_{ij} \right) + f_{\mathcal{ T}} \left( \mathcal{ T}_{ij} + \Theta_{ij}\right)  - \frac{1}{2} f g_{ij} +f_{Q} \left(2 Q^{\alpha \beta }_i P_{\alpha \beta j} -P_{i \alpha \beta} Q_{j}^{\alpha \beta}   \right)  
\end{equation}
\end{widetext}

\begin{equation}\label{4}
f_{i} = \frac{\partial f}{ \partial i}, \hspace{0.2in}  \mathcal{ T}_{ij} = -\frac{2}{\sqrt{-g}} \frac{\delta(\sqrt{-g}\mathcal{L}_{M})}{\delta g^{ij}}, \hspace{0.2in} \Theta_{ij} = g^{ij} \frac{\delta \mathcal{ T}_{ij}} {\delta g^{ij}}
\end{equation}
and $P^{\alpha}_{ij}$ is called superpotential and is defined as \cite{fqt}
\begin{equation}\label{5}
P^{\alpha}_{ij} = \frac{1}{4}\left[2 Q ^{\alpha}_{(i j)} - Q^{\alpha}_{i j} + Q ^{\alpha} g_{ij} - \delta^{\alpha} _{(i Q j)} - \tilde{Q}^{\alpha} g_{ij}\right] 
\end{equation}
where \begin{equation}\label{6}
Q_ \alpha = Q _{\alpha j} ^{j}, \hspace{0.25in} \tilde{Q}_{\alpha} = Q^{i}_{\alpha i}
\end{equation}
We now consider a flat FLRW spacetime of the form
\begin{equation}\label{7}
ds^{2} = -N^{2}(t)dt^{2}+ a^{2}(t)\sum_{i=1,2,3} \left(dx^{i} \right) ^{2}
\end{equation} 
where $a(t)$ represent the scale factor and the lapse function $N(t)=1$ for a flat spacetime. \\
Employing \eqref{7} in \eqref{3}, we finally obtain the modified Friedman equations with $N=1$  as \cite{fqt} 
\begin{equation}\label{8}
8 \pi \rho = -6 F H^{2} + \frac{f}{2} - \frac{2 \tilde{G}}{1 + \tilde{G}} \left(\dot{F}H + F \dot{H}\right)
\end{equation}
\begin{equation}\label{9}
8 \pi p =  6 F H^{2} - \frac{f}{2} +2 \left(\dot{F}H + F \dot{H}\right)
\end{equation}
where \begin{equation}
F = f_{Q}, \hspace{0.25in} \tilde{G} = \frac{f_{\mathcal{ T}}}{8 \pi}
\end{equation}
Combining equations \eqref{8} and \eqref{9}, we obtain the equation for Hubble parameter $H$ as
\begin{equation}\label{hub}
\dot{H}+\frac{\dot{F}}{F}H=\frac{4\pi}{F}\left(1 + \tilde{G}\right)(\rho+p)
\end{equation}

\section{$f(Q,\mathcal{ T})$ Baryogenesis}\label{sec:baryo}

According to cosmological observations such as CMB \cite{fg2} and BBN \cite{fg1}, the observed baryon to entropy ratio reads
\begin{equation}\label{13}
\frac{\eta_{B}}{s} \simeq 9 \times 10^{-11}
\end{equation} 
Sakharov reported three conditions for a net baryon asymmetry to occur through baryon number violation, C and CP violation and processes occurring outside of thermal equilibrium \cite{fg9}.\\
When the temperature $T$ falls below a critical value $T_{D}$ through the evolution of the Universe, the baryon to entropy ratio can be written as \cite{dav}
\begin{equation}\label{14}
\frac{\eta_{B}}{s} \simeq -\frac{15 g_{b}}{g_{*s}}\frac{\dot{R}}{M_{*}^{2} T_{D}}
\end{equation} 
where $g_{b}$ represent the total number of intrinsic degrees of freedom of the baryons, $g_{*s}$ represent the total number of degrees of freedom of the massless particles and the critical temperature $T_{D}$ is the temperature of the cosmos when all the interactions producing baryon asymmetry comes to a halt.\\
We shall presume that a thermal equilibrium prevails with energy density being associated with temperature $T$ as
\begin{equation}\label{15}
\rho(T) =\frac{\pi^{2}}{30} g_{*s} T^{4}
\end{equation}
Hence, for a CP violating interaction of \eqref{12}, the resulting baryon to entropy ratio in $f(Q,\mathcal{ T})$ gravity reads
\begin{equation}\label{16}
\frac{\eta_{B}}{s} \simeq - \frac{15 g_{b}}{4 \pi^{2} g_{*s}}\frac{(\dot{Q} + \dot{\mathcal{ T}})}{M_{*}^{2}T_{D}}
\end{equation}
We shall assume cosmological pressure and density obeys a barotropic equation of state of the form $p = (\gamma-1)\rho$, where $\gamma$ is a constant and $1\leq\gamma\leq2$.\\ 
Using the barotropic equation of state, from equations \eqref{8} and \eqref{hub} we obtain the matter density in general form as
\begin{equation}
\rho=\frac{f-12FH^2}{16\pi\left(1 +\gamma \tilde{G}\right)}.
\end{equation}\label{den}
For relativistic matter $p=\rho/3$ and hence $\mathcal{ T}=0$. Thus, the baryon to entropy ratio in $f(Q,\mathcal{ T})$ gravity for radiation dominated universe reduces to\\
\begin{equation}\label{17}
\frac{\eta_{B}}{s} \simeq - \frac{15 g_{b}}{4 \pi^{2} g_{*s}}\frac{\dot{Q}}{M_{*}^{2}T_{D}}
\end{equation}\\
We shall now compute the baryon-to-entropy ratio for a CP-violating interaction proportional to the nonmetricity $Q$ for two cases: First, with the Universe comprising predominantly of a perfect fluid with cosmological pressure $p$ and matter density $\rho$ following a barotropic equation of state and second, when the Universe is filled with a perfect fluid and the cosmic dynamics is governed by the $f(Q,\mathcal{ T})$ theory of gravity.

\subsection{The perfect fluid Case}

For a perfect fluid following barotropic equation of state, the Ricci scalar reads
\begin{equation}\label{18}
R = -8 \pi G (1-3(\gamma-1))\rho
\end{equation}
Thus, for a radiation dominated universe, $\gamma=4/3$ and hence $R=0$ which further implies \eqref{14} acquire a null value. Nonetheless, we shall show that when the baryon-to-entropy ratio is proportional to $\partial_{i}Q$ (Eq. \eqref{12}), the resultant baryon to entropy ratio is non-zero even for $\gamma=4/3$. Assuming a flat FLRW background with ($-$,$+$,$+$,$+$) metric signature and with the following expressions of scale factor $a(t)$ and energy density $\rho(t)$ in a radiation dominated universe as
\begin{equation}
a(t)=a_{0} t^{1/2}
\end{equation} 
\begin{equation}\label{29}
\rho = \rho_{0} a(t) ^{-4} = \rho_{0} t^{-2}
\end{equation}
the baryon to entropy ratio \eqref{17} reads\\
\begin{equation}\label{19}
\frac{\eta_{B}}{s} \simeq \frac{0.33 g_{b} \pi T_{D}^{5}}{8 M_{*}^{2} \rho_{0} \sqrt{\frac{\rho_{0}}{g_{*s}}}}
\end{equation}\\
where we have used $\dot{Q}=12H\dot{H}$ and the decoupling time $t_{D}$ is written in terms of critical temperature $T_{D}$ by equating \eqref{15} with \eqref{29} as

\begin{equation}\label{20}
t_{D}\simeq \sqrt{\frac{30 \rho_{0} }{\pi^{2}g_{*s}}}\left(\frac{1}{T_{D}} \right)^{2} 
\end{equation}\\

Substituting $g_{*s}=106$, $g_{b} \sim 1$, $\rho_{0}=3 \times 10^{26}GeV$, $T_{D}=2\times 10^{12} GeV$ and $M_{*}=2\times 10^{16} GeV$, the resultant baryon to entropy ratio reads
\begin{equation}
\frac{\eta_{B}}{s} \simeq 3.4 \times 10^{-11}
\end{equation}
which is close to the observational value \eqref{13}. Thus, the problem of baryogenesis can be resolved in Einstein's gravity if the CP-violating interactions are made proportional to the nonmetricity $Q$ instead of $R$.

\subsection{The perfect fluid with $f(Q,\mathcal{T})$ gravity case}

We shall now compute baryon to entropy ratio for the case when the Universe is filled with a perfect fluid and the evolution of the Universe is governed by the $f(Q,\mathcal{ T})$ theory of gravity. \\
We consider the $f(Q,\mathcal{ T})$ functional form to be \cite{fqt}
\begin{equation}\label{10}
f (Q,\mathcal{ T}) = \alpha Q ^{n+1} + \beta \mathcal{ T}
\end{equation}
where $\alpha$, $n$ and $\beta$ are model parameters. Substituting \eqref{10} in Eqs. \eqref{8} and \eqref{9}, the expression of Hubble parameter $H(t)$ and density $\rho (t)$ for this model reads \cite{fqt}
\begin{equation}\label{21}
H (t) = \frac{H_{0} (n+1)\left[16 \pi - \beta (\gamma - 4) \right] }{3 \gamma(\beta + 8 \pi) H_{0}(t-t_{0}) - (n+1)\left[\beta \gamma - 4 (\beta + 4 \pi) \right] }
\end{equation}
\begin{equation}\label{22}
\rho (t) = \frac{ \alpha 6 ^{(n+1)} (2 n +1) H (t)^{2(n+1)}}{\beta (\gamma -4) - 16 \pi}
\end{equation}
Equating \eqref{15} and \eqref{22}, the coupling time $t_{D}$ can be written as 

\begin{widetext}
\begin{equation}\label{23}
t_{D}=t_{0}+ \left[\frac{(1+n)\left( \frac{-1}{H^{0}}+5^{1/(2+2n)}\pi^{-1/(1+n)}\left(\frac{6^{-(2+n)}g_{*s}T_{D}^{4}(16\pi - \beta (\gamma-4))}{\alpha + 2 n \alpha} \right)^{-1/(2+2n)} \right)[16\pi - \beta (\gamma-4)] }{3(\beta+8 \pi)\gamma} \right]  
\end{equation}

where $H(t_{0})$ is the present value of the Hubble parameter.\\
Time derivative of Hubble parameter \eqref{21} reads

\begin{equation}\label{24}
\dot{H}=-\left[\frac{3H_{0}^{2}(1+n)(\beta+8\pi)[16\pi - \beta (\gamma-4)]\gamma}{(1+n)[16\pi - \beta (\gamma-4)]+3H_{0}^{2}(t-t_{0})(\beta+8 \pi)\gamma^{2}} \right] 
\end{equation}
Substituting \eqref{21}, \eqref{23} and \eqref{24} in \eqref{17}, the baryon to entropy ratio for $\gamma=4/3$ (radiation dominated universe) is given by 

\begin{equation}
\frac{\eta_{B}}{s} \simeq \frac{4}{3}\left[ \frac{9 \times 5^{[1-(3/2(1+n))]}g_{b}\pi^{[2+(3/(1+n))]}(\beta+8 \pi)\left(\frac{6^{(-2-n)}g_{*s}T_{D}^{4}[16\pi+ \frac{8 \beta}{3}]}{\alpha(1 + 2 n )} \right)^{[3/(2(1+n))]}}{4g_{*s} M_{*}^{2}(1+n)T_{D}[16\pi +\frac{8 \beta}{3}] }\right] 
\end{equation}
\end{widetext}

\begin{figure}[H]
\centering
\begin{subfigure}{.5\textwidth}
  \centering
  \includegraphics[width=8 cm]{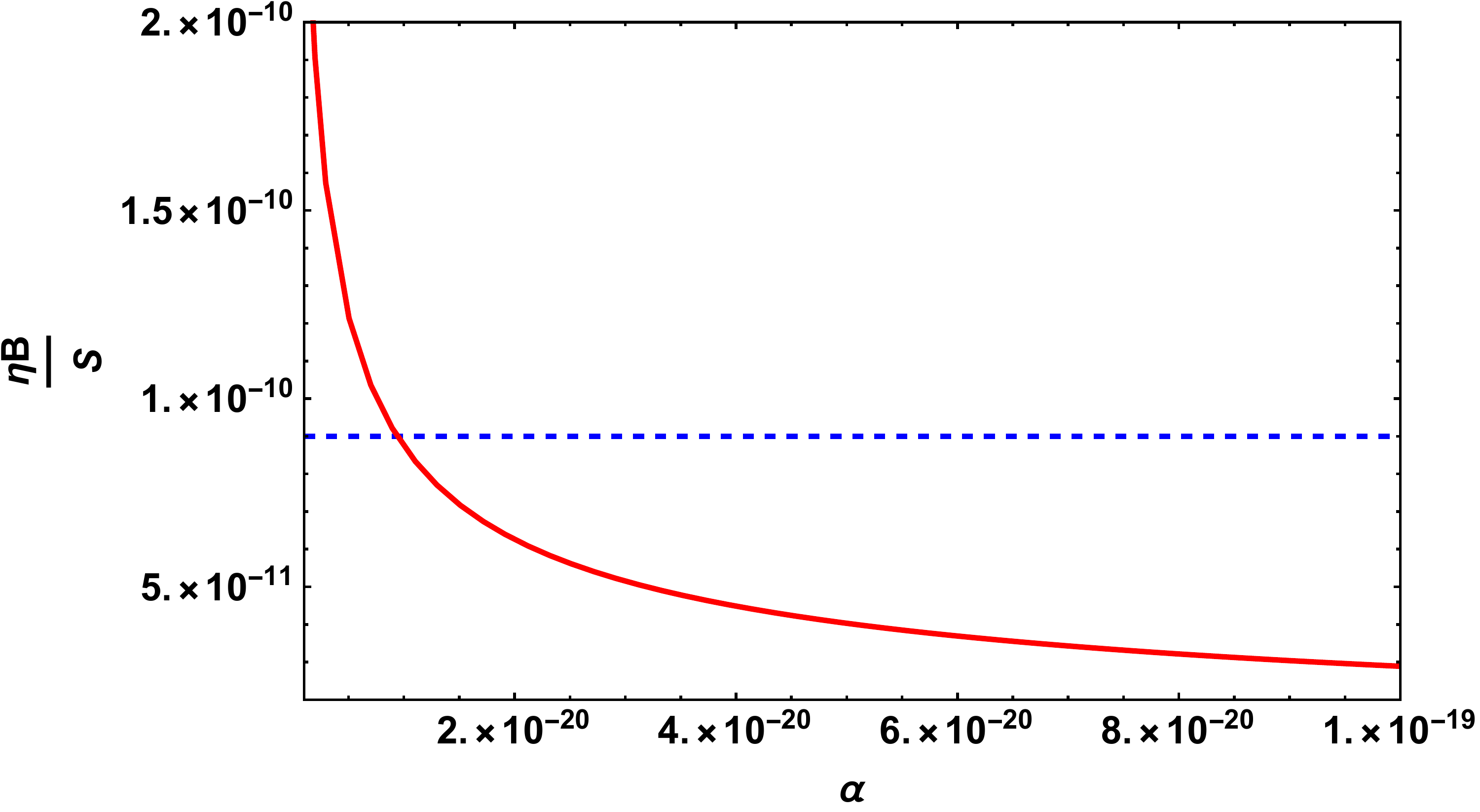}
   \caption{}
  \label{f1}
\end{subfigure}
\begin{subfigure}{.5\textwidth}
  \centering
  \includegraphics[width=8 cm]{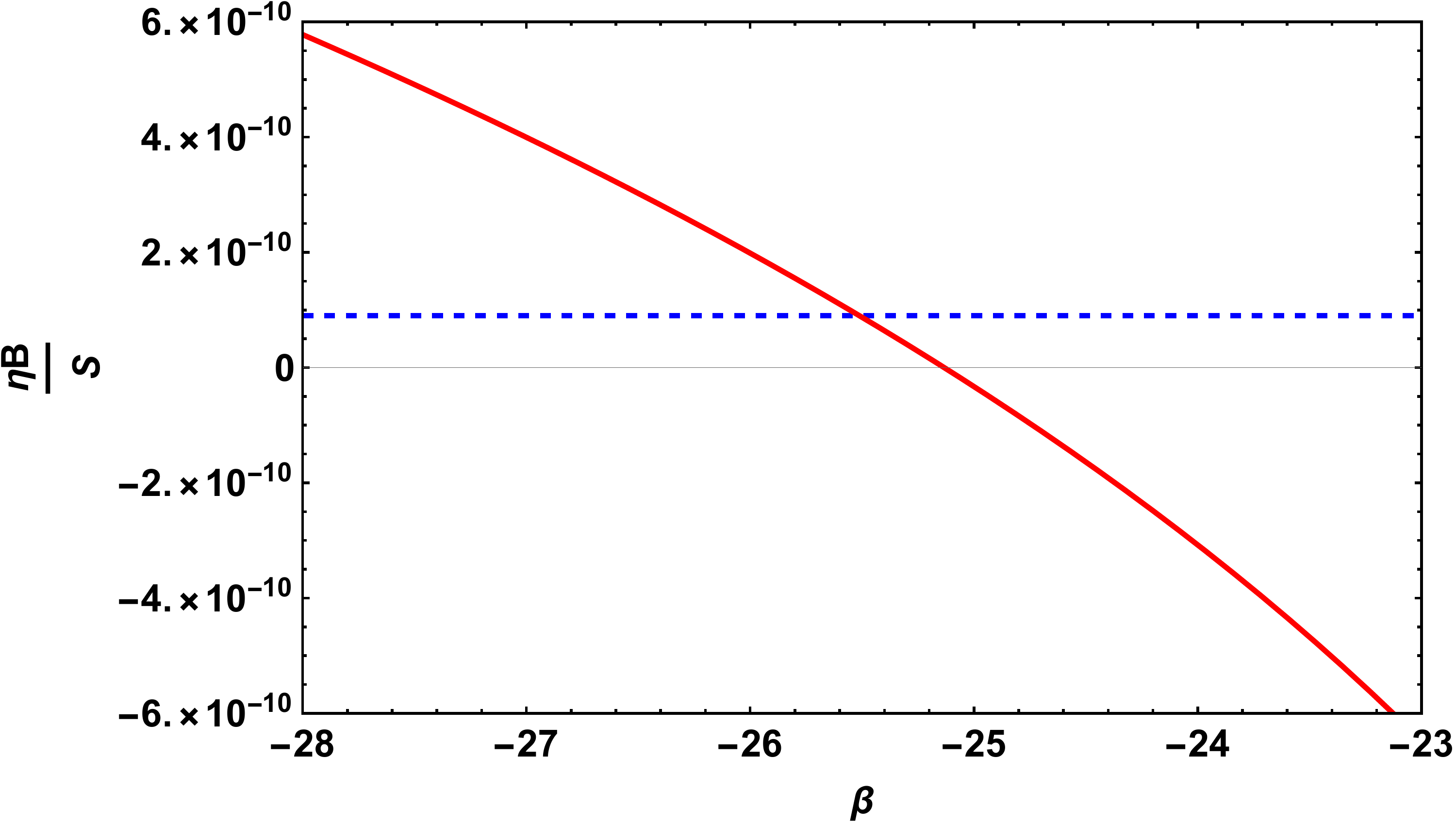}
   \caption{}
  \label{f2}
\end{subfigure}
\begin{subfigure}[C] {.5\textwidth}
  \centering
  \includegraphics[width=8 cm]{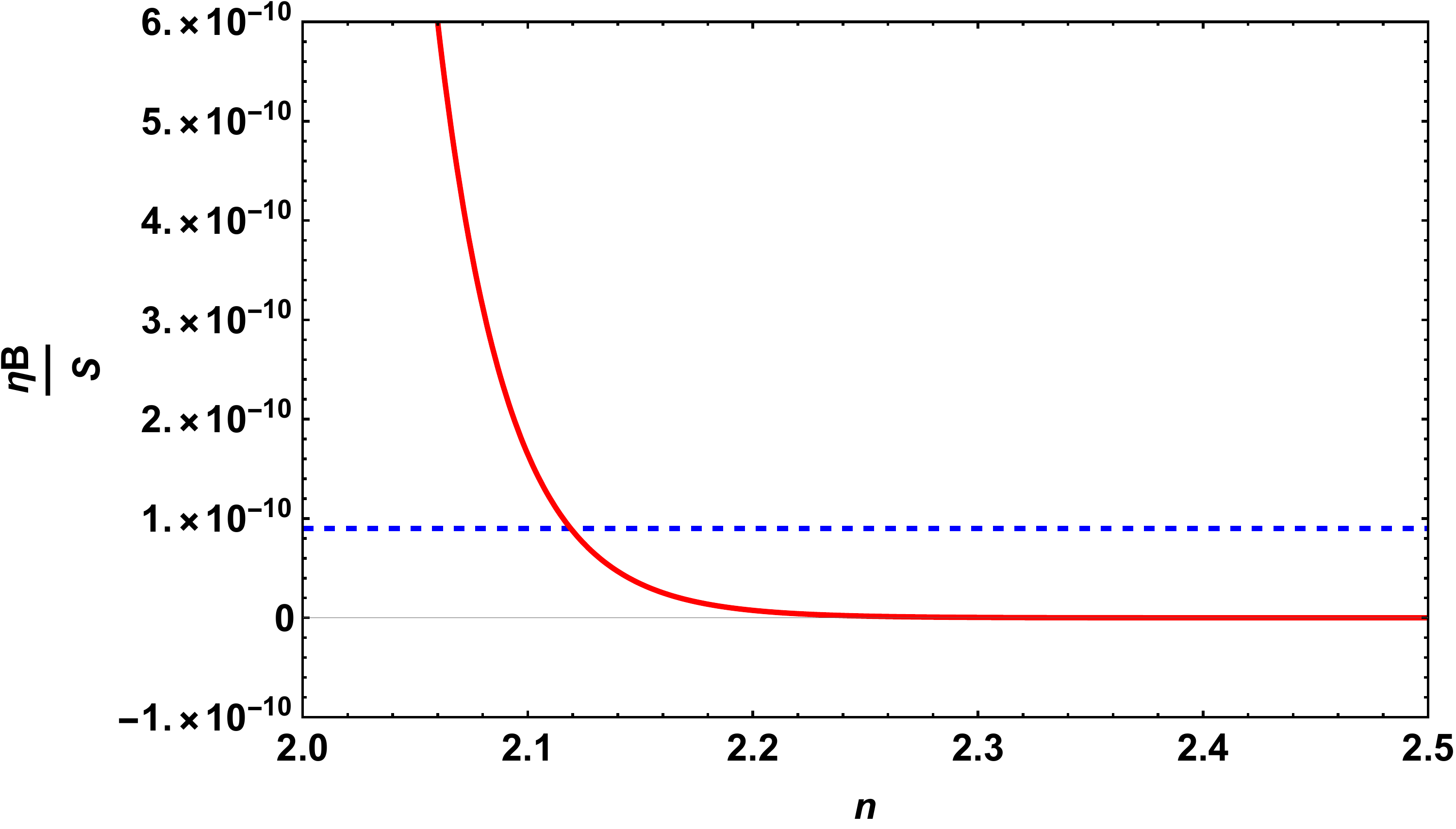}
   \caption{}
  \label{f3}
\end{subfigure}
\caption{\eqref{f1} shows $\eta_{B}/s$ as a function of $\alpha$, \eqref{f2} shows $\eta_{B}/s$ as a function of $\beta$ and \eqref{f3} shows $\eta_{B}/s$ as a function of $n$. We choose $g_{*s}=106$, $g_{b} \sim 1$ and $T_{D}\simeq M_{*}=2\times 10^{16} GeV$. }
\label{FIG1}
\end{figure}

By choosing $M_{*}$, $g_{*s}$, $g_{b}$, $T_{D}$ as before, $\alpha=10^{-20}$, $\beta = -8.1 \pi$ and $n=2.12$ the resultant baryon to entropy ratio reads $\sim6.17 \times 10^{-11}$ which is in excellent agreement with observations. In Fig. \eqref{f1}, \eqref{f2} and \eqref{f3} we show $\eta_{B}/s$ as a function of $\alpha$, $\beta$ and $n$ respectively.

Interestingly, in Fig. \eqref{f2}, the baryon to entropy ratio becomes negative for $\beta \lesssim -8 \pi$ which is unphysical as it implies an overabundance of antimatter over ordinary matter. Also note that for $n\gtrsim 2.2$, $\eta_{B}/s = 0$ which indicate no asymmetry between antimatter and matter and therefore not acceptable.
\subsubsection{Generalized Baryogenesis Interaction}
\begin{figure}[H]
\centering
\begin{subfigure}{.5\textwidth}
  \centering
  \includegraphics[width=7.5 cm]{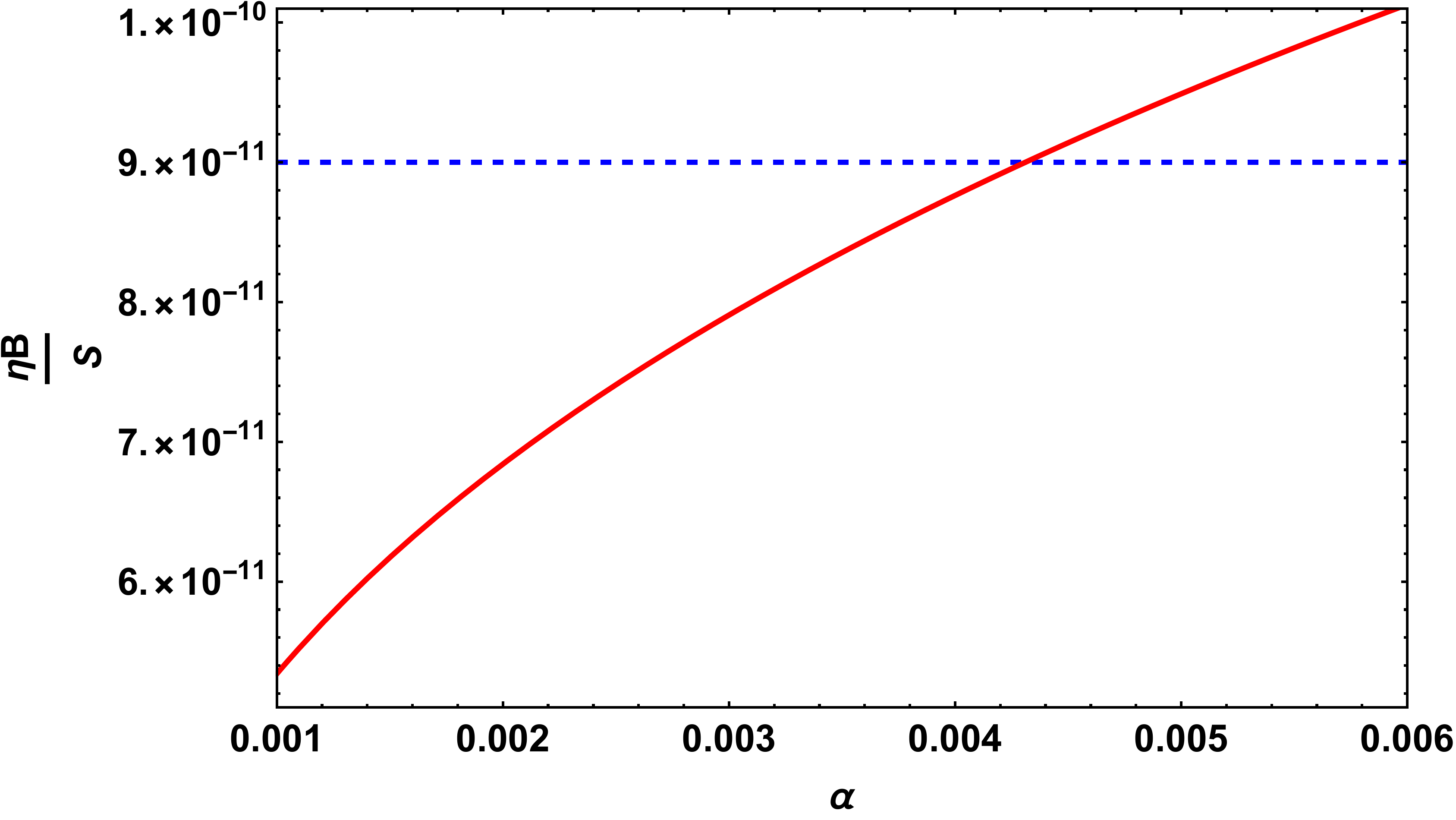}
   \caption{}
  \label{f4}
\end{subfigure}
\begin{subfigure}{.5\textwidth}
  \centering
  \includegraphics[width=7.5 cm]{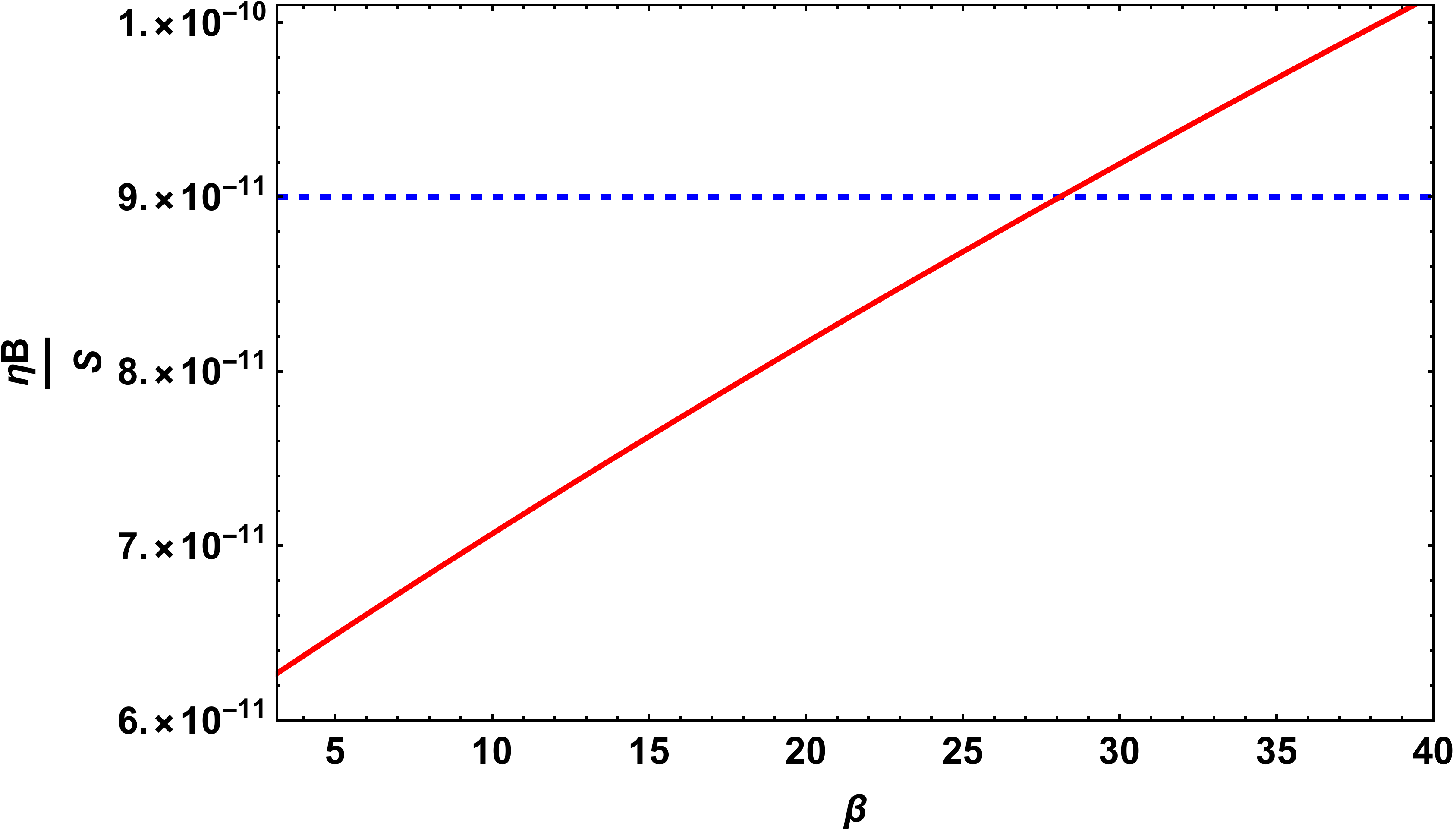}
   \caption{}
  \label{f5}
\end{subfigure}
\begin{subfigure}[C] {.5\textwidth}
  \centering
  \includegraphics[width=7.5 cm]{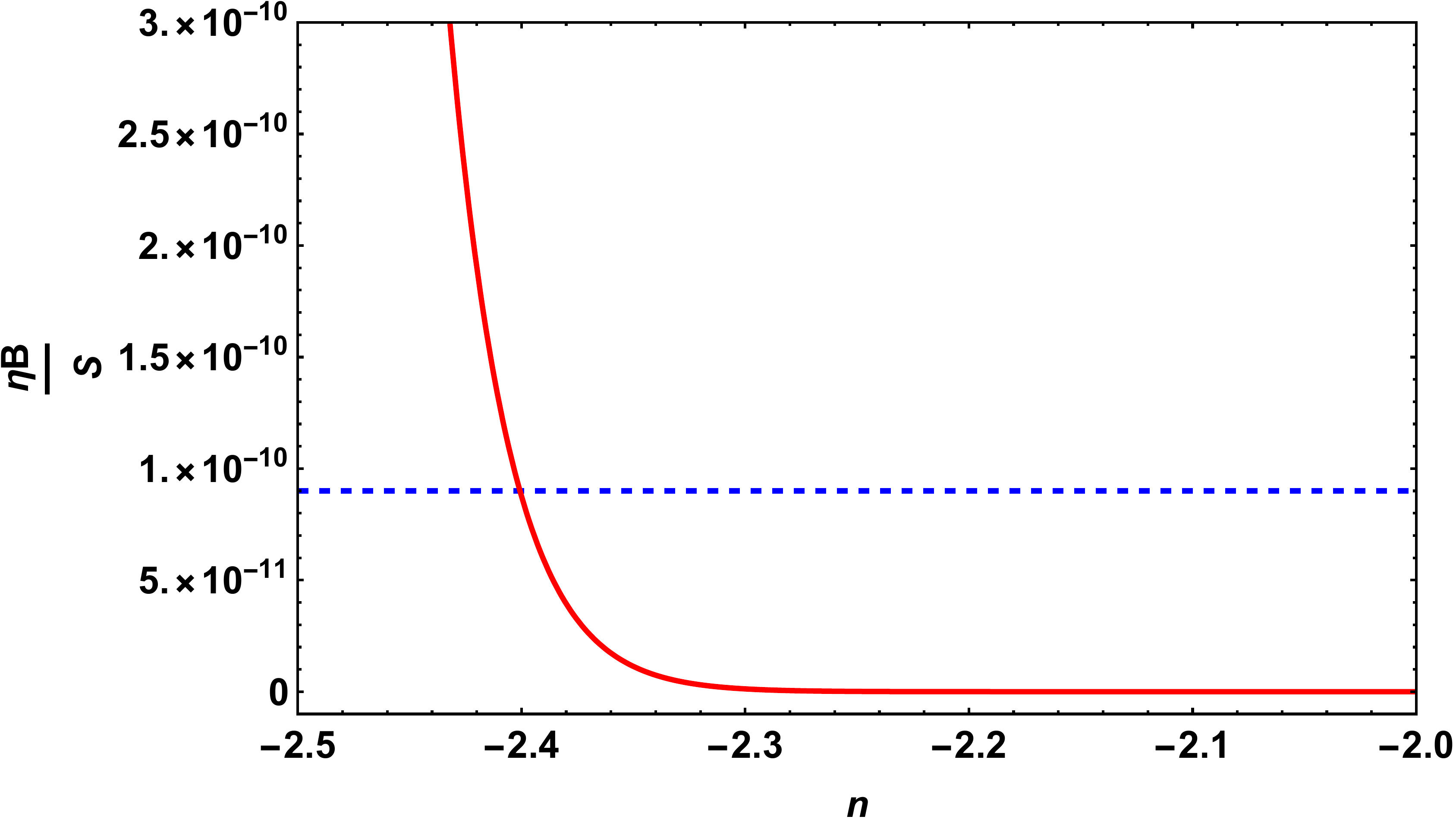}
   \caption{}
  \label{f6}
\end{subfigure}
\caption{\eqref{f4} shows $\eta_{B}/s$ as a function of $\alpha$, \eqref{f5} shows $\eta_{B}/s$ as a function of $\beta$ and \eqref{f6} shows $\eta_{B}/s$ as a function of $n$. We choose $g_{*s}=106$, $g_{b} \sim 1$ and $T_{D}\simeq M_{*}=2\times 10^{16} GeV$. }
\label{FIG2}
\end{figure}

We shall now define a more complete and generalized baryogenesis interaction proportional to $\partial_{i}f(Q,\mathcal{ T})$. The CP-violating interaction then reads
\begin{equation}\label{25}
\frac{1}{M_{*}^{2}}\int \sqrt{g}(\partial_{i}f(Q,\mathcal{ T}))J^{i} d^{4}x 
\end{equation}
For \eqref{25}, the resulting baryon to entropy ratio reads
\begin{equation}\label{26}
\frac{\eta_{B}}{s} \simeq - \frac{15 g_{b}}{4 \pi^{2} g_{*s}}\frac{(\dot{Q}f_{Q}+\dot{\mathcal{ T}}f_{\mathcal{ T}})}{M_{*}^{2}T_{D}}
\end{equation}
As discussed in the previous section that for a radiation dominated universe $\mathcal{ T}=0$, we finally obtain
\begin{equation}\label{27}
\frac{\eta_{B}}{s} \simeq - \frac{15 g_{b}}{4 \pi^{2} g_{*s}}\frac{\dot{Q}f_{Q}}{M_{*}^{2}T_{D}}
\end{equation}
Substituting \eqref{21}, \eqref{23} and \eqref{24} in \eqref{27}, the baryon to entropy ratio for $\gamma=4/3$ (radiation dominated universe) then reads
\begin{widetext}
\begin{equation}\label{28}
\frac{\eta_{B}}{s} \simeq  \frac{4}{3}\left[ \frac{2^{(2n)}3^{(3+2n)}5^{[1-(3/2(1+n))]}g_{b}\pi^{[2+(3/(1+n))]}\alpha(\beta+8 \pi)A^{[3/(2(1+n))]}\left(5^{[-1/2(1+n)]}\pi^{[1/(1+n)]}A \right)^{(2n)} }{g_{*s} M_{*}^{2}T_{D}[16\pi + \frac{8 \beta}{3}\beta ] }\right] 
\end{equation}
\end{widetext}
where 
\begin{equation}
A=\left(\frac{6^{-(2+n)}g_{*s}T_{D}^{4}[16\pi+ \frac{8 \beta}{3}]}{\alpha(1 + 2 n )} \right)
\end{equation}
Substituting $M_{*}$, $g_{*s}$, $g_{b}$, $T_{D}$ as before, $\alpha=0.004$, $\beta = 8.1 \pi$ and $n=-2.4$ the resultant baryon to entropy ratio reads $\eta_{B}/s \sim 8.7 \times 10^{-11}$ which is very close to observational constraints. In Fig. \eqref{f4}, \eqref{f5} and \eqref{f6} we show $\eta_{B}/s$ for the generalized baryogenesis interaction as a function of $\alpha$, $\beta$ and $n$ respectively.\\
Thus, the problem of baryogenesis can be resolved in $f(Q,\mathcal{T})$ gravity if the CP-violating interactions are made proportional to the nonmetricity $Q$ instead of $R$.

\newpage

\section{Conclusions}\label{sec:con}

The article presented a thorough investigation of gravitational baryogenesis interactions in the framework of $f(Q,\mathcal{ T})$ gravity where $Q$ denote the nonmetricity and $\mathcal{ T}$ the trace of the energy momentum tensor. For this type of modified gravity we find the baryon-to-entropy ratio to be proportional to $\dot{Q}$, since for a radiation dominated universe $\mathcal{ T}=0$. We ascertained the baryon-to-entropy ratio proportional to nonmetricity $Q$ two different scenarios, First, with the Universe comprising predominantly of a perfect fluid with cosmological pressure $p$ and matter density $\rho$ following a barotropic equation of state and second, when the Universe is filled with a perfect fluid and the cosmic dynamics is governed by the $f(Q,\mathcal{ T})$ theory of gravity. We choose the functional form of $f(Q,\mathcal{ T})$ gravity to be $f(Q,\mathcal{ T})=\alpha Q ^{n+1} + \beta \mathcal{ T}$, where $\alpha$, $\beta$ and $n$ are model parameters. For the perfect fluid case, the obtained baryon-to-entropy ratio $\eta_{B}/s \simeq 3.4 \times 10^{-11}$ while for the $f(Q,\mathcal{ T})$ gravity model we obtained $\eta_{B}/s \sim6.17 \times 10^{-11}$, both of which are in excellent agreement with observational value of $\simeq 9 \times 10^{-11}$. Next, for the $f(Q,\mathcal{ T})$ gravity model, we explored a more complete and generalized baryogenesis interaction proportional to $\dot{Q}f_{Q}$. For this baryogenesis interaction, we found the baryon-to-entropy ratio $\eta_{B}/s \sim 8.7 \times 10^{-11}$ which is very close to the observational value.

\textbf{Acknowledgments:} We are very much grateful to the honorable referee and the editor for the illuminating suggestions that have significantly improved our work in terms of research quality and presentation. SB thanks Biswajit Pandey for helpful discussions. PKS acknowledges CSIR, New Delhi, India for financial support to carry out the Research project [No.03(1454)/19/EMR-II Dt.02/08/2019].

\end{document}